\documentclass[twocolumn,dvipsnames]{fancyarticle}
\usepackage[utf8]{inputenc}
\usepackage{graphicx}
\usepackage[colorlinks,allcolors=cyan!70!black]{hyperref}
\usepackage{siunitx}

\title{An Open Source Mesh Generation Platform for Biophysical Modeling Using Realistic Cellular Geometries}
\runningtitle{An Open Source Mesh Generation Platform} 

\author[2,*]{Christopher T. Lee}
\author[2]{Justin G. Laughlin}
\author[3]{John B. Moody}
\author[1]{Rommie E. Amaro}
\author[1]{J. Andrew McCammon}
\author[3]{Michael J. Holst}
\author[2,*]{Padmini Rangamani}

\runningauthor{Lee et al.} 

\affil[1]{Department of Chemistry and Biochemistry, University of California, San Diego, La Jolla, CA, 92093 US}
\affil[2]{Department of Mechanical and Aerospace Engineering, University of California, San Diego, La Jolla, CA, 92093 USA}
\affil[3]{Department of Mathematics, University of California, San Diego, La Jolla, CA, 92093 USA}

\corrauthor[*]{ctlee@ucsd.edu, prangamani@ucsd.edu}

\usepackage[color=green!20]{todonotes}

\usepackage[capitalise]{cleveref}
\usepackage{xspace}

\newcommand{\casc}{\texttt{CASC}\xspace}
\newcommand{\gamer}{\texttt{GAMer}\xspace}
\newcommand{\gamertwo}{\texttt{GAMer} \texttt{2}\xspace}
\newcommand{\cmake}{\texttt{CMake}\xspace}

\newcommand{\tetgen}{\texttt{TetGen}\xspace}
\newcommand{\blender}{\texttt{Blender}\xspace}
\newcommand{\pygamer}{\texttt{PyGAMer}\xspace}
\newcommand{\blendgamer}{\texttt{BlendGAMer}\xspace}
\newcommand{\pybind}{\texttt{PyBind} \texttt{11}\xspace}
\newcommand{\fenics}{\texttt{FEniCS}\xspace}

\usepackage[acronym, nonumberlist, nopostdot, nogroupskip, numberedsection=false]{glossaries}
\setacronymstyle{long-short}
\makeglossaries

\newacronym{et}{ET}{Electron Tomography}
\newacronym{lst}{LST}{Local Structure Tensor}
\newacronym{fem}{FEM}{Finite Element Method}

\newacronym{api}{API}{Application Programming Interface}
\newacronym{pde}{PDE}{Partial Differential Equation}

\begin{document}

\begin{frontmatter}

\begin{abstract}
Advances in imaging methods such as electron microscopy, tomography, and other modalities are enabling high-resolution reconstructions of cellular and organelle geometries.
Such advances pave the way for using these geometries for biophysical and mathematical modeling once these data can be represented as a geometric mesh, which, when carefully conditioned, enables the discretization and solution of partial differential equations.
In this study, we outline the steps for a na\"{i}ve user to approach \gamertwo, a mesh generation code written in C++ designed to convert structural datasets to realistic geometric meshes, while preserving the underlying shapes.
We present two example cases, 1) mesh generation at the subcellular scale as informed by electron tomography, and 2) meshing a protein with structure from x-ray crystallography.
We further demonstrate that the meshes generated by \gamer are suitable for use with numerical methods.
Together, this collection of libraries and tools simplifies the process of constructing realistic geometric meshes from structural biology data.
\end{abstract}

\begin{sigstatement}
As biophysical structure determination methods improve, the rate of new structural data is increasing.
New methods that allow the interpretation, analysis, and reuse of such structural information will thus take on commensurate importance.
In particular, geometric meshes, such as those commonly used in graphics and mathematics, can enable a myriad of mathematical analysis.
In this work, we describe \gamertwo, a mesh generation library designed for biological datasets.
Using \gamertwo and associated tools \pygamer and \blendgamer, biologists can robustly generate computer and algorithm friendly geometric mesh representations informed by structural biology data.
We expect that \gamertwo will be a valuable tool to bring realistic geometries to biophysical models.
\end{sigstatement}
\end{frontmatter}

\begin{figure*}[!h]
    \centering
    \includegraphics[width=\textwidth]{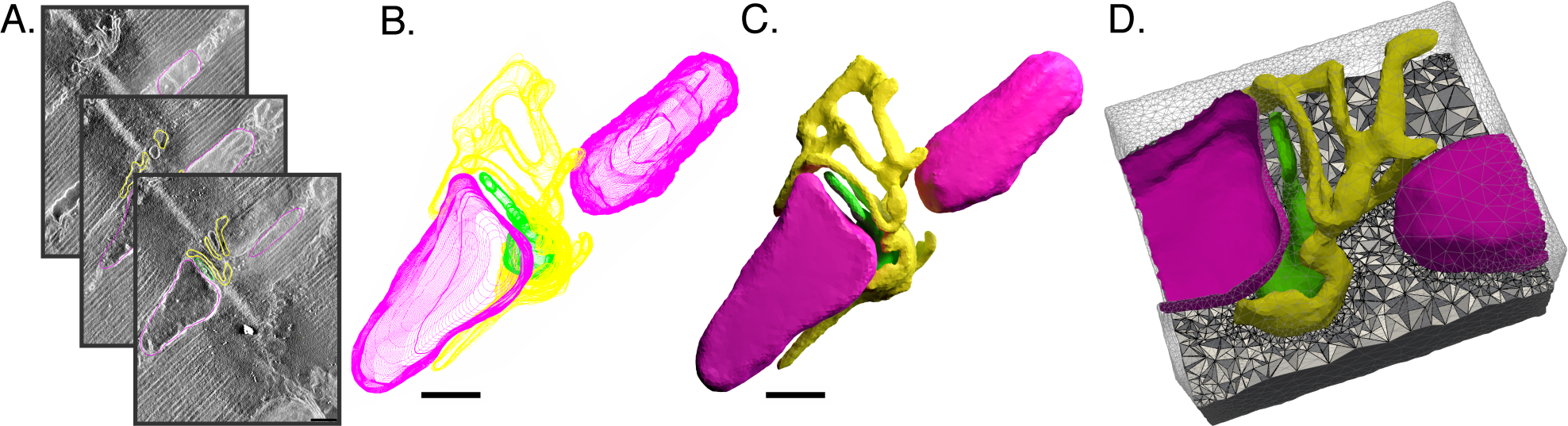}
    \caption{
        Example workflow using \gamer to construct a tetrahedral domain suitable for use with Finite Element simulations.
        A) Segmented Electron Tomogram of a murine cardiac Calcium Release Unit (CRU) from the Cell Image Library entry 3603.
        B) Stacks of contours from traced model.
        C) Conditioned surface mesh of the model.
        D) Tetrahedralized domain which can be used for simulating cytosolic diffusion.
        Note that we have inverted the tetrahedralized domain to represent the free space surrounding the CRU geometry.
        Scale bars are \SI{200}{\nano\meter}.
    }
    \label{fig:workflow}
\end{figure*}

\section*{Introduction}
The use of \glspl{pde} in mathematical modeling of cellular phenomena is becoming increasing common, particularly, for problems that include electrostatics, reaction-diffusion, fluid dynamics, and continuum mechanics.
Solutions to these equations using idealized geometries have provided insight into how cell shape can affect signaling \cite{Rangamani2013,Bell1017,cugno_geometric_2019,ohadi_computational_2019}, and how blood flows in vessels \cite{simvascular}.

On the other hand, in order to gain better insight into how cellular geometry can affect the dynamics of these mechanochemical processes, using realistic geometries is necessary.
Already, freely available tools such as \texttt{Virtual Cell} \cite{Loew2001} and \texttt{CellOrganizer} \cite{Murphy2012} have paved the way for using realistic cellular geometries in simulations.
With increasing availability of high-resolution images of cellular ultrastructure, including the size and shape of organelles, and the curvature of the various cellular membranes, there is a need for computational tools and algorithms that can enable us to use these data as the geometry or domain of interest and conduct simulations using numerical methods \cite{Xu2017}.

For most relevant geometries, it is impossible to obtain analytical solutions for \glspl{pde}; this necessitates the use of numerical methods to provide an approximate solution.
These numerical methods are based on \emph{discretization} (approximating the \gls{pde} with a discrete algebraic system) combined with \emph{solvers} (typically iterative methods that converge to the solution to the algebraic system).
The first step usually requires the generation of a geometric mesh over which the problem can be discretized using techniques such as finite difference, finite volume, finite element, or other methods to build the algebraic system that approximates the \gls{pde}.
The numerical approximation to the \gls{pde} is then produced by solving the resulting linear or nonlinear algebraic equations using an appropriate fast solver.
One of the computational challenges associated with generating meshes of biological datasets is the presence of highly irregular surfaces with curvatures at the nanometer or \AA ngstrom length scales.
Although many tools from the graphics community exist to generate meshes for visualization, these poor quality meshes, when used to solve a \gls{pde}, can both destroy the quality of the discretization as an approximation to the \gls{pde}, and also produce algebraic systems that are badly conditioned and difficult to solve efficiently or accurately with iterative solvers.

While it is possible to design discretizations of \gls{pde} problems on surfaces using finite volume methods or other techniques, we prefer the \gls{fem} here for a number of reasons.
To begin with, the \gls{fem} first arose in the 1960s in the engineering community as a response to the poor performance of existing discretization techniques for \glspl{pde} involving shells and other complex physical shapes.
In addition, methods such as the \gls{fem} that are built on basis function expansion provide a natural framework both for treating highly nonlinear problems, and for discretizing multiple \glspl{pde} that couple together to form a larger multi-physics system.
Lastly, the \gls{fem} framework is quite general, and can in fact be shown to reproduce finite volume, spectral, and other discretizations through appropriate choices of basis and test functions.

One challenge preventing the routine use of experimental ultrastructural data with \gls{pde}-based mathematical modeling is the difficulty associated in generating a discretization, or commonly a mesh, which accurately represents the structures of interest.
Building upon existing meshing codes, such as \tetgen \cite{Si2015}, \texttt{NetGen} \cite{Schoberl1997}, \texttt{TetWild} \cite{Hu2018}, \texttt{MeshLab} \cite{meshlab}, \texttt{Gmsh} \cite{Geuzaine2009}, \texttt{CGAL} \cite{cgal}, along with commercial codes such as \texttt{ANSYS Meshing} among others, we describe the development of a meshing framework, the Geometry-preserving Adaptive MeshER software version 2 (\gamertwo), which is designed specifically for biological mesh generation.
This code has been completely rewritten from version 1, which was previously described by Yu \textit{et al.} \cite{Yu2008b,Yu2008}, Gao \textit{et al.} \cite{GYH11,GYH12}, and \cite{ChHo10a}.
\gamertwo features the original \gamer algorithms but with significantly improved ease of use, distributability, and maintainability.
We have also developed a new Python \gls{api}: \pygamer, and streamlined the \gamer \blender add-on: \blendgamer.
The complete explanation of the mathematics and underlying algorithms are available in \cite{Yu2008b,Yu2008,GYH11,GYH12,ChHo10a,gamer2}.
In what follows, we provide an overview of the \gamer mesh generation capabilities for the general biophysicist.

\section*{Methods}

At its core, \gamer is a mesh generation and conditioning library which uses concepts from the graphics and engineering literature.
It can be used to produce high quality surface and volume meshes from several types of input: 1) PDB/PQR file (\AA ngstrom--nanometer), 2) Volumetric dataset (\AA ngstrom--meters), or 3) Initial surface mesh (\AA ngstrom--meters).
To enable \gls{fem}-based applications, \gamer also has utilities to support boundary marking.
Tetrahedral meshes of a domain can be constructed in \gamer through the use of functionality provided by \tetgen \cite{tetgen}.
Surface or volume meshes can be output to several common formats for use with other tools such as \fenics \cite{LoggMardalEtAl2012a,AlnaesBlechta2015a}, \texttt{ParaView} \cite{Ahrens2005}, \texttt{MCell} \cite{mcell}, and \texttt{FFEA} \cite{ffea} among others.
We note that although \gamer is designed primarily with \gls{fem}-based applications in mind, conditioned meshes of realistic geometries can also be used for geometric analysis such as the estimation of curvatures, volumes, and surface areas \cite{Lee534479}.
Conditioned meshes can also be used in other tools such as \texttt{MCell} \cite{mcell} and 3D printing.
Example tutorials of using \gamertwo to generate surface and volume meshes of a protein structure (PDB ID: 2JHO) and a subcellular scene of a calcium release unit (CRU) from a murine cardiac myocyte imaged using \gls{et} (Cell Image Library: 3603 \cite{cru2004}), \cref{fig:workflow}, are provided in the documentation\footnote{\url{https://gamer.readthedocs.io/en/latest/tutorial.html}} and described in this report.
Here we will summarize the key implementation steps and refer the interested reader to \cite{Lee534479} for technical details of the implementation.

\subsection*{GAMer 2 Development}

One of the limitations of prior versions of \gamer is the inability to robustly represent meshes of arbitrary manifoldness and topology.
This limitation prevented the safe application of \gamer to non-manifold meshes, which often produced segmentation faults or other undefined behaviors.
To ameliorate this, in \gamertwo we replaced the underlying mesh representation to use the Colored Abstract Simplicial Complex (\casc) data structure \cite{casc_Lee2019}.
\casc keeps careful track of the mesh topology and therefore makes it trivial to track the manifoldness of a given mesh object.
By eliminating the need for code to handle encounters with non-manifold elements, the code base is significantly reduced and segmentation faults are eliminated.
Another benefit of using \casc is that it can represent both surface meshes (2D simplices embedded in 3D) and volume meshes (3D simplices embedded in 3D), allowing users to interact with both surface and volume meshes using an identical \gls{api}.
This simplification contributes to the long-term maintainability of the \gamertwo code.

In the development of \gamertwo, we have also migrated to use the cross-platform \cmake build toolchain, and away from the previous GNU build system \texttt{Autotools}.
Using \cmake, \gamertwo can now be compiled and run on major operating systems including Linux, and Mac OS--along with Windows which was previously unsupported.
Detailed installation instructions are provided online\footnote{\url{https://gamer.readthedocs.io/en/latest/install.html}}.
We note that the Windows binary can be built using \texttt{Microsoft Visual Studio} and does not require the installation of Unix-like environments such as \texttt{Cygwin}.
By supporting compilation of \gamertwo using native and preferred tools, this improves binary compatibility with other libraries and simplifies distribution.

\subsubsection*{Collaborative Workflow}

To further improve code availability and to encourage community collaboration, the \gamer code is now hosted on Github\footnote{\url{https://github.com/ctlee/gamer}}.
In this environment, users can file issues to report bugs or ask questions.
Users are also encouraged to contribute to the code by submitting pull requests.
All pull requests are subject to rigorous continuous-integration testing, using both Travis-CI (Linux and Mac OS) and Appveyor (Windows), to ensure code compatibility across a wide range of operating systems, compilers, and compiler versions prior to integration with the main deployment branches.
Along with source control, \gamertwo also implements git tagging based semantic versioning to track the software version.
Compiled code can report the source version which aids in reporting and debugging.
This strict versioning introduces improved ability for both users and developers to track code provenance.

\subsection*{Implementation of a New PyGAMer API}

In addition to the complete redesign of the core library, the corresponding Python interface, now called \pygamer, is generated using \pybind \cite{pybind11} instead of \texttt{SWIG} \cite{swig}.
\pybind was designed to expose C++ data types to Python and vice-versa while minimizing boilerplate code by capitalizing upon the capabilities of the C++ compiler.
One of the benefits of this approach is the ability to bind complex template types, which are extensively used in \casc, enabling users to develop Python script to interact with various elements of the mesh and call C++ methods.
Another benefit of using \pybind is its support for embedding Python docstrings, which enable straightforward documentation of \pygamer using popular Python tools such as \texttt{Sphinx}.
As a result, documentation for \gamertwo and \pygamer is now automatically generated and hosted online\footnote{\url{https://gamer.readthedocs.io}}.

Using \texttt{scikit-build}, which connects \texttt{setuptools} and \cmake, installation of \pygamer in any Python environment can be achieved using \texttt{pip install pygamer}.
Versions of \texttt{pip}$\geq$10.0 will automatically download and resolve buildtime and runtime dependencies, compile, and install the \pygamer Python extension module.
Users of older \texttt{pip} versions need to install any missing buildtime dependencies beforehand.
Instructions on how to install these dependencies is provided in both the online documentation and demonstrated in Supplementary Movie S1.

\begin{figure}[!h]
    \centering
    \includegraphics[width=\columnwidth]{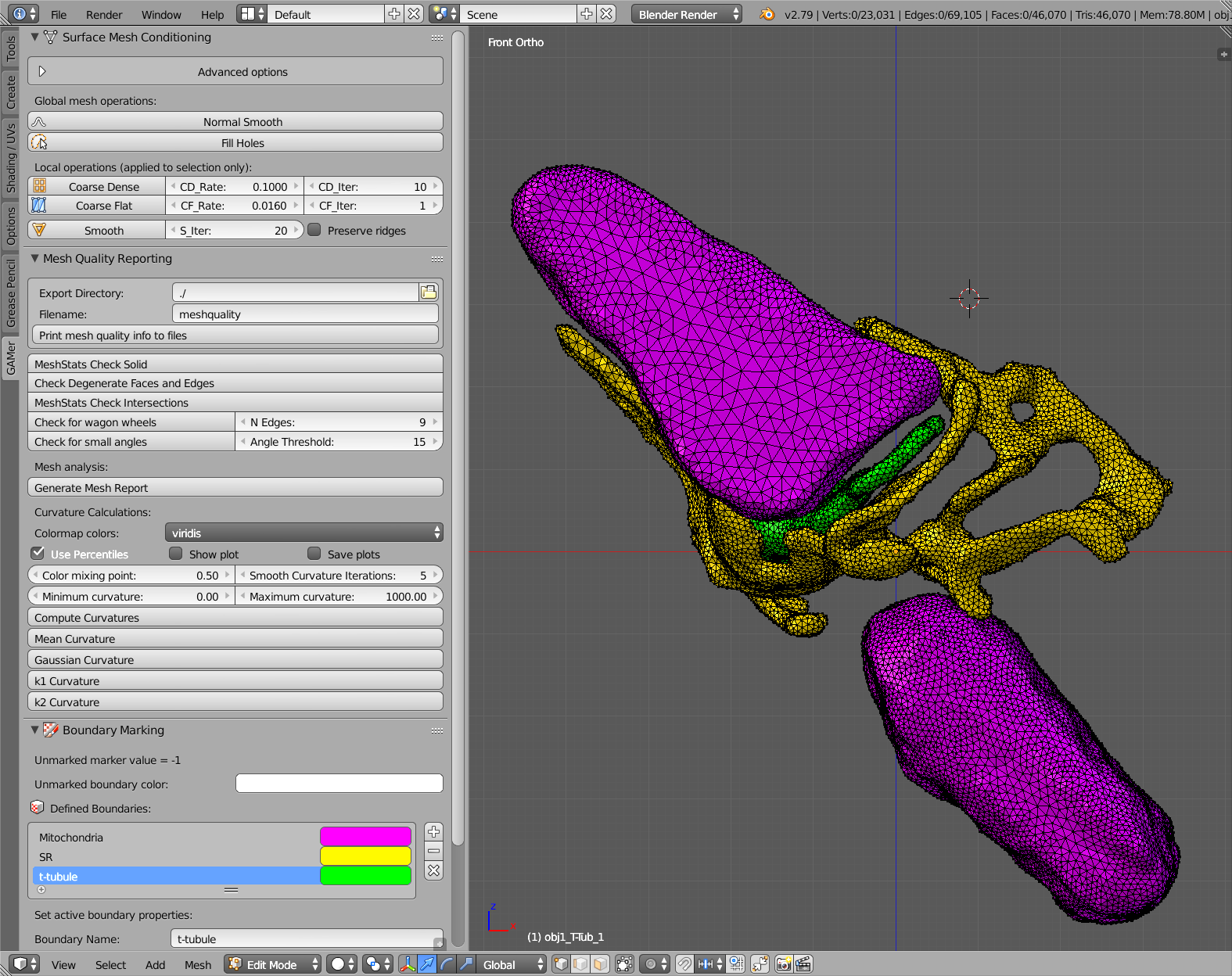}
    \caption{
        Screenshot of the \blendgamer toolshelf menu in 3D modeling software \blender.
        The user can call \gamer mesh conditioning, analysis, boundary marking, and tetrahedralization functions by clicking buttons and adjusting settings in the toolshelf.
        Shown on the right is a conditioned surface mesh of the calcium release unit model.
    }
    \label{fig:blendgamer}
\end{figure}

\subsection*{BlendGAMer Development}

In order to support interactive modeling with graphical feedback, we have also developed a \gamer addon for \blender \cite{blender}, \blendgamer, which has also been rewritten to use \pygamer.
In addition to this update, the user interface has been redesigned to be user-friendly as shown in \cref{fig:blendgamer}.
The boundary marking capability of \blendgamer now uses \blender attribute layers instead of lists of values.
Many features now have corresponding toggles in the interface, for example, the number of neighborhood rings to consider when computing the \gls{lst}.
Several mesh conditioning operations have also been updated to operate only upon selected vertices.
This provides users with the flexibility to refine local portions of the mesh as desired.
There is also a new Mesh Analysis panel that contains several helpful features for analyzing the quality of a mesh and includes curvature estimation.
Based upon the newly implemented semantic versioning, \blendgamer can now track the version of metadata that is stored in a given file.
Using this information, \blendgamer can perform automatic metadata migration from version to version as improved schemes for metadata storage are created.

\subsection*{Modeling Diffusion in the CRU Geometry}

To demonstrate the use of the mesh generated from \gamertwo with the \gls{fem}, we model the diffusion of a molecule with concentration $u$ in the realistic geometry of a CRU, \cref{fig:crudiffusion}.
In the volume, the dynamics of $u$ are given by
\begin{align}
\frac{\partial u}{\partial t} &= D\nabla^2 u - \frac{u}{\tau} ~~\text{in}~~ \Omega, \\
u(\mathbf{x},t=0) &= 0,
\end{align}
where $D$ is the diffusion coefficient, $\tau$ is a decay constant and represents reactions that consume $u$, $\Omega$ is the cytosolic domain, and $t$ is time.
We define the following boundary conditions:
\begin{align}
D(n \cdot \nabla u) &= J_\text{in} \enspace\text{on}\enspace \partial\Omega_\text{t-tubule}, \\
D(n \cdot \nabla u) &= 0 \enspace\text{on}\enspace \partial\Omega_\text{other}, \\
\partial\Omega &= \partial\Omega_\text{t-tubule}  \cup \partial\Omega_\text{other},
\end{align}
where $J_\text{in}$ is the inward flux on the t-tubule membrane ($\partial \Omega_\text{t-tubule}$).
No flux boundary conditions are applied to all other boundaries.
The following system is solved using \fenics \cite{AlnaesBlechta2015a,LoggMardalEtAl2012a} and visualized using \texttt{ParaView} \cite{Ahrens2005}.
We note that the boundaries $\partial \Omega_\text{t-tubule}$ and $\partial \Omega_\text{other}$ are differentiated by markers applied using \blendgamer.

\section*{Results and Discussion}

We demonstrate that \gamer is capable of generating high quality surface and volume simplex meshes of geometries as informed by structural biology datasets.
The incorporated Python library, \pygamer, and \blender add-on, \blendgamer, enable users to prototype or interact with meshes as they are conditioned.
Collectively these tools facilitate mathematical modeling of biological systems using realistic geometries.

\begin{figure}[!ht]
    \centering
    \includegraphics[width=\columnwidth]{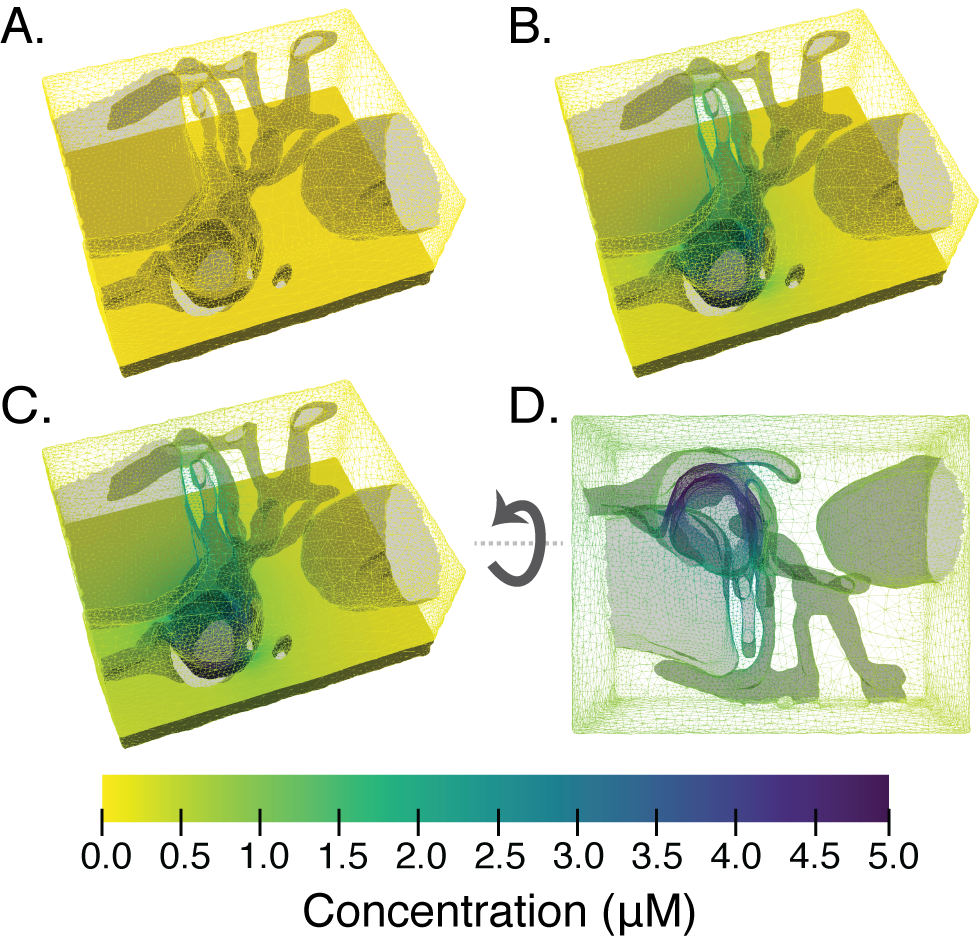}
    \caption{Snapshots of molecular diffusion in the calcium release unit geometry.
    A) initial condition; B) \SI{200}{\micro\second}; C) \SI{400}{\micro\second}; D) \SI{5000}{\micro\second}.
    The molecules can be trapped in locally confined regions leading to microdomains with locally increased concentrations.
    }
    \label{fig:crudiffusion}
\end{figure}

The \gamer workflow has been previously applied in several works \cite{Yu2008c,Cheng2010,Cheng2012,Hake2012,Kekenes-Huskey2012,BromerE2410,Lee534479}.
Several example applications are also described in the online \gamertwo documentation.
As an example, the steps to go from \gls{et} data to simulation quality mesh is shown in \cref{fig:workflow}.
In this example, a segmented \gls{et} dataset (\cref{fig:workflow}~A) featuring a murine CRU is retrieved from the Cell Image Library, shown in \cref{fig:workflow}~B.
We note that this is the same starting geometry previously used by Hake et al. \cite{Hake2012}.
From the model contours, we generated a preliminary mesh that was conditioned using \blendgamer to produce \cref{fig:workflow}~C.
To model the cytosolic space surrounding the CRU, \blender Boolean mesh operations were used to invert the geometry followed by tetrahedralization (\cref{fig:workflow}~D).
The mesh is now sufficiently high quality and suitable for use with \gls{fem}-based simulations as shown in \cref{fig:crudiffusion}.

\begin{figure}[!ht]
    \centering
    \includegraphics[width=\columnwidth]{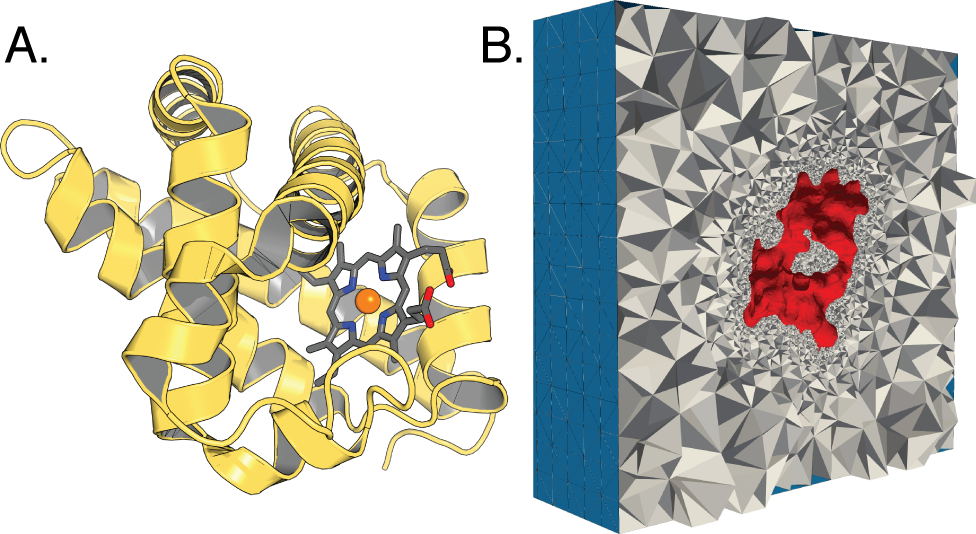}
    \caption{Example demonstrating the meshing of protein myoglobin (PDB ID: 2JHO).
    A) Rendered cartoon representation with heme and iron shown as sticks.
    B) Tetrahedralization of the space excluding the protein volume.
    Red denotes faces marked as protein and blue denotes faces on the boundary of the enclosing cube.
    }
    \label{fig:protein}
\end{figure}

In \cref{fig:protein}, we demonstrate the mesh generation capabilities of \gamer from atomistic protein structural data such as those available from the Protein Data Bank.
A volume dataset representing the approximate occupied space of all atoms is generated by applying a Gaussian kernel over the atomic positions.
An isosurface of the dataset can then be meshed using the marching cubes algorithm \cite{Lorensen}.
While \gamer includes a basic table of atom types to assign radii, more sophisticated atomic radius assignment tools (e.g., \texttt{PDB2PQR} \cite{Dolinsky2004}) can be used and radius information passed via the PQR file format.

\section*{Conclusion}

The realism of biophysical models can be enhanced by incorporation realistic geometries from structural biology at various length scales.
We have demonstrated the applicability of \gamertwo at two different length scales and suggest that tools such as \gamertwo brings the community closer to realizing the goal of conducting physics-based simulations in realistic cellular geometries by simplifying the mesh generation process.
We believe that, moving forward, \gamertwo can serve as an integrative platform for meshing biological mesh generation.
Furthermore, our design strategies in implementing \gamertwo encourage community collaboration, an important aspect of tool building for systems and computational biology.

\section*{Software Availability}

\gamer is licensed under GNU Lesser General Public
License (LGPL), version 2.1 or later.
Full documentation and examples are available at the project home page, gamer.readthedocs.io, and development is hosted on GitHub at \url{http://github.com/ctlee/gamer}.
The latest release v2.0.5 is archived on Zenodo (doi:10.5281/zenodo.2340294).

\section*{Author Contributions}

C.T.L., J.G.L., J.B.M., and M.J.H. developed the software;
C.T.L. drafted the article;
C.T.L., J.G.L., J.B.M., J.A.M., R.E.A., M.J.H., and P.R. edited the article;
and all authors read and approved the final article.

\section*{Acknowledgments}

CTL, JBM, REA, JAM, and MJH are supported in part by the National Institutes of Health under grant number P41-GM103426.
CTL, and JAM are also supported by the NIH under R01-GM31749.
CTL also acknowledges support from the NIH Molecular Biophysics Training Grant T32-GM008326 and a Hartwell Foundation Postdoctoral Fellowship.
MJH was supported in part by the National Science Foundation under awards DMS-CM1620366 and DMS-FRG1262982.
PR was supported by the Air Force Office of Scientific Research (AFOSR) Multidisciplinary University Research Initiative (MURI) FA9550-18-1-0051 and
JGL was supported by a fellowship from the UCSD Center for Transscale Structural Biology and Biophysics/Virtual Molecular Cell Consortium.

REA is a cofounder and scientific advisor of, and has equity interest in, Actavalon, Inc.

\bibliography{minimal.bib}

\begin{thebibliography}{40}
\providecommand{\url}[1]{\texttt{#1}}
\providecommand{\urlprefix}{ }

\bibitem[Rangamani et~al.()Rangamani, Lipshtat, Azeloglu, Calizo, Hu, Ghassemi,
  Hone, Scarlata, Neves, and Iyengar]{Rangamani2013}
Rangamani, P., A.~Lipshtat, E.~U. Azeloglu, R.~C. Calizo, M.~Hu, S.~Ghassemi,
  J.~Hone, S.~Scarlata, S.~R. Neves, and R.~Iyengar.
\newblock Decoding Information in Cell Shape 154:1356--1369.
\newblock
  \urlprefix\url{http://www.sciencedirect.com/science/article/pii/S0092867413010209}.

\bibitem[Bell et~al.()Bell, Bartol, Sejnowski, and Rangamani]{Bell1017}
Bell, M., T.~Bartol, T.~Sejnowski, and P.~Rangamani.
\newblock Dendritic spine geometry and spine apparatus organization govern the
  spatiotemporal dynamics of calcium 151:1017--1034.
\newblock \urlprefix\url{http://jgp.rupress.org/content/151/8/1017}.

\bibitem[Cugno et~al.()Cugno, Bartol, Sejnowski, Iyengar, and
  Rangamani]{cugno_geometric_2019}
Cugno, A., T.~M. Bartol, T.~J. Sejnowski, R.~Iyengar, and P.~Rangamani.
\newblock Geometric principles of second messenger dynamics in dendritic spines
  9:1--18.
\newblock \urlprefix\url{https://www.nature.com/articles/s41598-019-48028-0}.

\bibitem[Ohadi et~al.()Ohadi, Schmitt, Calabrese, Halpain, Zhang, and
  Rangamani]{ohadi_computational_2019}
Ohadi, D., D.~L. Schmitt, B.~Calabrese, S.~Halpain, J.~Zhang, and P.~Rangamani.
\newblock Computational {Modeling} {Reveals} {Frequency} {Modulation} of
  {Calcium}-{cAMP}/{PKA} {Pathway} in {Dendritic} {Spines}
  \urlprefix\url{http://www.sciencedirect.com/science/article/pii/S0006349519308318}.

\bibitem[Updegrove et~al.()Updegrove, Wilson, Merkow, Lan, Marsden, and
  Shadden]{simvascular}
Updegrove, A., N.~M. Wilson, J.~Merkow, H.~Lan, A.~L. Marsden, and S.~C.
  Shadden.
\newblock {SimVascular}: An Open Source Pipeline for Cardiovascular Simulation
  45:525--541.

\bibitem[Loew and Schaff()]{Loew2001}
Loew, L.~M., and J.~C. Schaff.
\newblock {The Virtual Cell: A software environment for computational cell
  biology}.

\bibitem[Murphy()]{Murphy2012}
Murphy, R.~F.
\newblock {CellOrganizer: Image-Derived Models of Subcellular Organization and
  Protein Distribution}.
\newblock \emph{In} Methods Cell Biol.

\bibitem[Xu et~al.()Xu, Hayworth, Lu, Grob, Hassan, Garc{í}a-Cerd{á}n,
  Niyogi, Nogales, Weinberg, and Hess]{Xu2017}
Xu, C.~S., K.~J. Hayworth, Z.~Lu, P.~Grob, A.~M. Hassan, J.~G.
  Garc{í}a-Cerd{á}n, K.~K. Niyogi, E.~Nogales, R.~J. Weinberg, and H.~F.
  Hess.
\newblock {Enhanced FIB-SEM systems for large-volume 3D imaging} 6.
\newblock \urlprefix\url{https://elifesciences.org/articles/25916}.

\bibitem[Si({\natexlab{a}})]{Si2015}
Si, H.
\newblock {TetGen, a Delaunay-Based Quality Tetrahedral Mesh Generator}
  41:1--36.
\newblock \urlprefix\url{http://dl.acm.org/citation.cfm?id=2732672.2629697}.

\bibitem[Sch{ö}berl()]{Schoberl1997}
Sch{ö}berl, J.
\newblock {An advancing front 2D/3D-mesh generator based on abstract rules} .

\bibitem[Hu et~al.()Hu, Zhou, Gao, Jacobson, Zorin, and Panozzo]{Hu2018}
Hu, Y., Q.~Zhou, X.~Gao, A.~Jacobson, D.~Zorin, and D.~Panozzo.
\newblock {Tetrahedral meshing in the wild} 37:1--14.
\newblock \urlprefix\url{http://dl.acm.org/citation.cfm?doid=3197517.3201353}.

\bibitem[Cignoni et~al.()Cignoni, Callieri, Corsini, Dellepiane, Ganovelli, and
  Ranzuglia]{meshlab}
Cignoni, P., M.~Callieri, M.~Corsini, M.~Dellepiane, F.~Ganovelli, and
  G.~Ranzuglia.
\newblock {MeshLab: an Open-Source Mesh Processing Tool}.
\newblock \emph{In} V.~Scarano, R.~{De Chiara}, and U.~Erra, editors,
  Eurographics Ital. Chapter Conf. The Eurographics Association.

\bibitem[Geuzaine and Remacle()]{Geuzaine2009}
Geuzaine, C., and J.-F. Remacle.
\newblock {Gmsh: A 3-D finite element mesh generator with built-in pre- and
  post-processing facilities} 79:1309--1331.
\newblock \urlprefix\url{http://doi.wiley.com/10.1002/nme.2579}.

\bibitem[cga()]{cgal}
{CGAL, Computational Geometry Algorithms Library}.
\newblock \urlprefix\url{http://www.cgal.org}.

\bibitem[Yu et~al.({\natexlab{a}})Yu, Holst, Cheng, and McCammon]{Yu2008b}
Yu, Z., M.~J. Holst, Y.~Cheng, and J.~McCammon.
\newblock {Feature-Preserving Adaptive Mesh Generation for Molecular Shape
  Modeling and Simulation} 26:1370--1380.

\bibitem[Yu et~al.({\natexlab{b}})Yu, Holst, and {Andrew McCammon}]{Yu2008}
Yu, Z., M.~J. Holst, and J.~{Andrew McCammon}.
\newblock {High-Fidelity Geometric Modeling for Biomedical Applications}
  44:715--723.

\bibitem[Gao et~al.({\natexlab{a}})Gao, Yu, and Holst]{GYH11}
Gao, Z., Z.~Yu, and M.~Holst.
\newblock Quality Tetrahedral Mesh Smoothing via Boundary-Optimized {Delaunay}
  Triangulation 29:707--721.

\bibitem[Gao et~al.({\natexlab{b}})Gao, Yu, and Holst]{GYH12}
Gao, Z., Z.~Yu, and M.~Holst.
\newblock Feature-Preserving Surface Mesh Smoothing via Suboptimal {Delaunay}
  Triangulation 75:23--38.

\bibitem[Chen and Holst()]{ChHo10a}
Chen, L., and M.~Holst.
\newblock Efficient Mesh Optimization Schemes Based on Optimal {Delaunay}
  Triangulations 200:967--984.

\bibitem[Lee et~al.({\natexlab{a}})Lee, Moody, Laughlin, and Holst]{gamer2}
Lee, C.~T., J.~B. Moody, J.~G. Laughlin, and M.~J. Holst.
\newblock {GAMer 2.0 Software}.
\newblock \urlprefix\url{https://github.com/ctlee/gamer}.

\bibitem[Si({\natexlab{b}})]{tetgen}
Si, H.
\newblock TetGen, a Delaunay-Based Quality Tetrahedral Mesh Generator
  41:11:1--11:36.
\newblock \urlprefix\url{http://doi.acm.org/10.1145/2629697}.

\bibitem[Logg et~al.()Logg, Mardal, Wells, et~al.]{LoggMardalEtAl2012a}
Logg, A., K.-A. Mardal, G.~N. Wells, et~al.
\newblock Automated Solution of Differential Equations by the Finite Element
  Method.
\newblock Springer.

\bibitem[Aln{æ}s et~al.()Aln{æ}s, Blechta, Hake, Johansson, Kehlet, Logg,
  Richardson, Ring, Rognes, and Wells]{AlnaesBlechta2015a}
Aln{æ}s, M.~S., J.~Blechta, J.~Hake, A.~Johansson, B.~Kehlet, A.~Logg,
  C.~Richardson, J.~Ring, M.~E. Rognes, and G.~N. Wells.
\newblock The FEniCS Project Version 1.5 3.

\bibitem[Ahrens et~al.()Ahrens, Geveci, and Law]{Ahrens2005}
Ahrens, J., B.~Geveci, and C.~Law.
\newblock {ParaView: An end-user tool for large-data visualization}.
\newblock \emph{In} Vis. Handb.

\bibitem[Stiles and Bartol()]{mcell}
Stiles, J.~R., and T.~M. Bartol.
\newblock Monte {C}arlo methods for simulating realistic synaptic
  microphysiology using {MC}ell.
\newblock \emph{In} E.~D. Schutter, editor, Computational Neuroscience:
  Realistic Modeling for Experimentalists, CRC Press, 87--127.

\bibitem[Solernou et~al.()Solernou, Hanson, Richardson, Welch, Read, Harlen,
  and Harris]{ffea}
Solernou, A., B.~S. Hanson, R.~A. Richardson, R.~Welch, D.~J. Read, O.~G.
  Harlen, and S.~A. Harris.
\newblock Fluctuating Finite Element Analysis (FFEA): A continuum mechanics
  software tool for mesoscale simulation of biomolecules 14:1--29.
\newblock \urlprefix\url{https://doi.org/10.1371/journal.pcbi.1005897}.

\bibitem[Lee et~al.({\natexlab{b}})Lee, Laughlin, de~La~Beaumelle, Amaro,
  McCammon, Ramamoorthi, Holst, and Rangamani]{Lee534479}
Lee, C.~T., J.~G. Laughlin, N.~A. de~La~Beaumelle, R.~E. Amaro, J.~A. McCammon,
  R.~Ramamoorthi, M.~J. Holst, and P.~Rangamani.
\newblock GAMer 2: A system for 3D mesh processing of cellular electron
  micrographs
  \urlprefix\url{https://www.biorxiv.org/content/early/2019/07/23/534479}.

\bibitem[Hoshijima et~al.()Hoshijima, Hayashi, Thor, Terada, Martone, and
  Ellisman]{cru2004}
Hoshijima, M., T.~Hayashi, A.~Thor, M.~Terada, M.~Martone, and M.~Ellisman.
\newblock {CCDB:3603, MUS MUSCULUS, T-tubules, sarcoplasmic reticulum,
  myocyte}.
\newblock CIL. Dataset.

\bibitem[Lee et~al.({\natexlab{c}})Lee, Moody, Amaro, Mccammon, and
  Holst]{casc_Lee2019}
Lee, C.~T., J.~B. Moody, R.~E. Amaro, J.~A. Mccammon, and M.~J. Holst.
\newblock {The Implementation of the Colored Abstract Simplicial Complex and
  Its Application to Mesh Generation} 45:1--20.
\newblock \urlprefix\url{http://dl.acm.org/citation.cfm?doid=3349340.3321515}.

\bibitem[Jakob et~al.()Jakob, Rhinelander, and Moldovan]{pybind11}
Jakob, W., J.~Rhinelander, and D.~Moldovan.
\newblock pybind11 -- Seamless operability between C++11 and Python.
\newblock Https://github.com/pybind/pybind11.

\bibitem[Beazley()]{swig}
Beazley, D.~M.
\newblock SWIG: An Easy to Use Tool for Integrating Scripting Languages with C
  and C++.
\newblock \emph{In} Proceedings of the 4th Conference on USENIX Tcl/Tk
  Workshop, 1996 - Volume 4. USENIX Association, TCLTK'96, 15--15.
\newblock \urlprefix\url{http://dl.acm.org/citation.cfm?id=1267498.1267513}.

\bibitem[Community()]{blender}
Community, B.~O.
\newblock Blender - a 3D modelling and rendering package.
\newblock Blender Foundation.
\newblock \urlprefix\url{http://www.blender.org}.

\bibitem[Yu et~al.({\natexlab{c}})Yu, Holst, Hayashi, Bajaj, Ellisman,
  McCammon, and Hoshijima]{Yu2008c}
Yu, Z., M.~J. Holst, T.~Hayashi, C.~L. Bajaj, M.~H. Ellisman, J.~A. McCammon,
  and M.~Hoshijima.
\newblock {Three-dimensional geometric modeling of membrane-bound organelles in
  ventricular myocytes: Bridging the gap between microscopic imaging and
  mathematical simulation} 164:304--313.
\newblock
  \urlprefix\url{http://linkinghub.elsevier.com/retrieve/pii/S1047847708002281}.

\bibitem[Cheng et~al.({\natexlab{a}})Cheng, Yu, Hoshijima, Holst, McCulloch,
  McCammon, and Michailova]{Cheng2010}
Cheng, Y., Z.~Yu, M.~Hoshijima, M.~J. Holst, A.~D. McCulloch, J.~A. McCammon,
  and A.~P. Michailova.
\newblock {Numerical Analysis of Ca2+ Signaling in Rat Ventricular Myocytes
  with Realistic Transverse-Axial Tubular Geometry and Inhibited Sarcoplasmic
  Reticulum} 6:e1000972.
\newblock \urlprefix\url{http://dx.plos.org/10.1371/journal.pcbi.1000972}.

\bibitem[Cheng et~al.({\natexlab{b}})Cheng, Kekenes-Huskey, Hake, Holst,
  McCammon, and Michailova]{Cheng2012}
Cheng, Y., P.~Kekenes-Huskey, J.~E. Hake, M.~J. Holst, J.~A. McCammon, and
  A.~P. Michailova.
\newblock {Multi-scale continuum modeling of biological processes: from
  molecular electro-diffusion to sub-cellular signaling transduction} 5:015002.
\newblock
  \urlprefix\url{https://iopscience.iop.org/article/10.1088/1749-4699/5/1/015002}.

\bibitem[Hake et~al.()Hake, Edwards, Yu, Kekenes-Huskey, Michailova, McCammon,
  Holst, Hoshijima, and McCulloch]{Hake2012}
Hake, J., A.~G. Edwards, Z.~Yu, P.~M. Kekenes-Huskey, A.~P. Michailova, J.~A.
  McCammon, M.~J. Holst, M.~Hoshijima, and A.~D. McCulloch.
\newblock {Modelling cardiac calcium sparks in a three-dimensional
  reconstruction of a calcium release unit} 590:4403--4422.
\newblock \urlprefix\url{http://doi.wiley.com/10.1113/jphysiol.2012.227926}.

\bibitem[Kekenes-Huskey et~al.()Kekenes-Huskey, Cheng, Hake, Sachse, Bridge,
  Holst, McCammon, McCulloch, and Michailova]{Kekenes-Huskey2012}
Kekenes-Huskey, P.~M., Y.~Cheng, J.~E. Hake, F.~B. Sachse, J.~H. Bridge, M.~J.
  Holst, J.~A. McCammon, A.~D. McCulloch, and A.~P. Michailova.
\newblock {Modeling effects of L-type ca(2+) current and na(+)-ca(2+) exchanger
  on ca(2+) trigger flux in rabbit myocytes with realistic T-tubule
  geometries.} 3:351.
\newblock
  \urlprefix\url{http://journal.frontiersin.org/article/10.3389/fphys.2012.00351/abstract}.

\bibitem[Bromer et~al.()Bromer, Bartol, Bowden, Hubbard, Hanka, Gonzalez,
  Kuwajima, Mendenhall, Parker, Abraham, Sejnowski, and Harris]{BromerE2410}
Bromer, C., T.~M. Bartol, J.~B. Bowden, D.~D. Hubbard, D.~C. Hanka, P.~V.
  Gonzalez, M.~Kuwajima, J.~M. Mendenhall, P.~H. Parker, W.~C. Abraham, T.~J.
  Sejnowski, and K.~M. Harris.
\newblock Long-term potentiation expands information content of hippocampal
  dentate gyrus synapses 115:E2410--E2418.
\newblock \urlprefix\url{https://www.pnas.org/content/115/10/E2410}.

\bibitem[Lorensen and Cline()]{Lorensen}
Lorensen, W.~E., and H.~E. Cline.
\newblock Marching Cubes: A High Resolution 3D Surface Construction Algorithm
  21:163--169.
\newblock \urlprefix\url{http://doi.acm.org/10.1145/37402.37422}.

\bibitem[Dolinsky et~al.()Dolinsky, Nielsen, McCammon, and Baker]{Dolinsky2004}
Dolinsky, T.~J., J.~E. Nielsen, J.~A. McCammon, and N.~A. Baker.
\newblock {PDB2PQR: an automated pipeline for the setup of Poisson-Boltzmann
  electrostatics calculations} 32:W665--W667.
\newblock
  \urlprefix\url{https://academic.oup.com/nar/article-lookup/doi/10.1093/nar/gkh381}.

\end{thebibliography}

\end{document}